# A HIGH PHASE ADVANCE DAMPED AND DETUNED STRUCTURE FOR THE MAIN LINACS OF CLIC

V.F. Khan[†*], A. D'Elia[†*‡], A. Grudiev[‡], R.M. Jones[†*], W. Wuensch[‡]
[†]School of Physics and Astronomy, The University of Manchester, Manchester, U.K.
[*]The Cockcroft Institute of Accelerator Science and Technology, Daresbury, U.K.
[‡]CERN, Geneva, Switzerland

*Abstract*

The main accelerating structures for the CLIC are designed to operate at an average accelerating gradient of 100 MV/m. The accelerating frequency has been optimised to 11.994 GHz with a phase advance of $2\pi/3$ [1] of the main accelerating mode. The moderately damped and detuned structure (DDS) design [2-3] is being studied as an alternative to the strongly damped WDS design [1]. Both these designs are based on the nominal accelerating phase advance. Here we explore high phase advance (HPA) structures in which the group velocity of the rf fields is reduced compared to that of standard ($2\pi/3$) structures. The electrical breakdown strongly depends on the fundamental mode group velocity. Hence it is expected that electrical breakdown is less likely to occur in the HPA structures. We report on a study of both the fundamental and dipole modes in a CLIC_DDS_HPA structure, designed to operate at $5\pi/6$ phase advance per cell. Higher order dipole modes in both the standard and HPA structures are also studied.

## INTRODUCTION

The motivation behind studying the DDS for CLIC is to provide an alternative design to the baseline heavily damped design of Q~10 [1]. The DDS designs rely on moderate damping of the higher order modes (HOMs). As part of DDS study for CLIC, a test structure known as CLIC_DDS_A [4] has been designed and is presently under mechanical fabrication. CLIC_DDS_A is designed for high power testing (71 MW) and it operates at a $2\pi/3$ phase advance per cell. The power absorbed during an rf breakdown in the cavity is proportional to the square of the fundamental mode group velocity [5]. This provides the motivation for reducing the overall group velocity. The group velocity is a strong function of the iris radius of the accelerating cells; reducing the iris radius will reduce the group velocity. However, at the same time, the short range wakefield, which is inversely proportional to the 4th power of the iris radius [6], will also increase. For a given iris radius, as the synchronous phase approaches the π mode, the group velocity approaches zero. Hence, if we choose a high synchronous phase for accelerating the particles, then the group velocity can be suitably managed without imposing intense short range wakefields. By taking advantage of the comprehensive study of the HPA structures designed and tested for the Next Linear Collider (NLC) [2-3], we rely on the same phase advance per cell of $5\pi/6$ to study the CLIC_DDS_HPA structures.

As the transverse HOMs (dipole in particular) can severely dilute the beam emittance we study the first six dipole bands in both structures ($2\pi/3$ and $5\pi/6$). In these multi-band simulations, the first dipole band dominates all other bands. Hence the focus of our work is on damping the lowest dipole band. In order to properly suppress the wakefields excited by the lowest dipole band, within the prescribed beam dynamics limit [1], it is necessary to interleave eight structures (each of which contains 24 cells).

The DDS_HPA structure presented in this paper satisfies both the rf breakdown and beam dynamics criterion [1]. With a similar rf-to-beam efficiency, the surface fields in this structure have been minimised and the wakefields are well-damped compared to the DDS_A structure.

## ACCELERATING MODE

Both the accelerating (fundamental) mode and the higher order mode rf properties of the accelerating structures are particularly sensitive to the aperture radius of the cells. In order to compare these rf properties of the standard and HPA structure, we retain the standard structure iris dimensions (which range from 4mm to 2.5mm) [7]. We also retain the elliptical cavity wall shape, in order to minimise the surface magnetic field (H-field) [4], [7]. Changing the phase advance per cell to $5\pi/6$ increases the cell length by ~25%. In simulations of DDS_A it was observed that the maximum field enhancement was in the vicinity of the manifold coupling slots. Hence, in order to minimise the perturbation of the fundamental mode, we optimised the coupling slots of the manifold such that they are now at a distance of $R_c = 8.5$ mm from the electrical centre of the cavity.

The group velocity of the fundamental mode is also a function of the thickness of the iris. Changing the iris thickness also affects the shunt impedance of the structure which in itself modifies the beam loading properties. We investigated a range of iris thicknesses to optimise the surface fields, rf-to-beam efficiency and lowest dipole coupling to the attached manifolds. The fundamental mode rf properties of the HPA structure, for a range of iris thicknesses are illustrated in Fig. 1. It can be observed that a very low group velocity does not necessarily reduce the overall surface fields, due to the increase in the unloaded gradient in the lower half of the structure. In this optimisation, an iris thickness ranging from 3.2 mm to 2.8 mm has been chosen. A comparison of the fundamental mode rf properties of the DDS_A structure with the HPA structure is presented in Table 1.

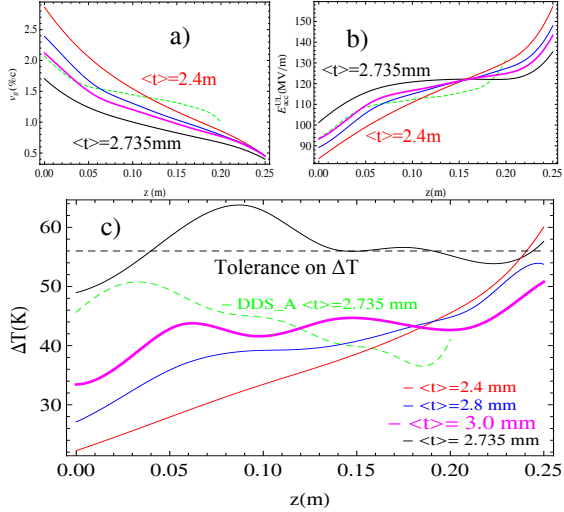

Figure 1: RF properties of the fundamental mode of the HPA for a range of iris thicknesses. a) Group velocity, b) Unloaded accelerating gradient for $4.2 \times 10^9$ particles per bunch, c) Pulsed temperature rise for a pulse length of 269 ns.

Table 1: Comparison of the Fundamental Mode Rf Properties of Standard and HPA Structure

| RF parameters | DDS_A | DDS_HPA42 | DDS_HPA32 |
|---|---|---|---|
| Iris thickness (In/Out)[mm] | 4/1.74 | 3.2/2.8 | 3.2/2.8 |
| Bunch population ($n_b$)[$10^9$] | 4.2 | 4.2 | 3.2 |
| $v_g/c$ (In/Out) [%] | 2.07 / 1.0 | 2.1 / 0.45 | 2.1 / 0.45 |
| $P_{in}$ [MW] | 71 | 68.2 | 63.6 |
| $E_{acc}$(Load./UnL.) [ MV/m] | 105 / 132 | 93 / 143 | 90 / 138 |
| $\Delta T_{sur}$[$^\circ$K] | 51 | 51 | 48 |
| $E_{sur}$ [MV/m] | 220 | 234 | 225 |
| $S_c$ [W/$\mu m^2$] | 6.75 | 5.9 | 5.5 |
| RF-beam effi. ($\eta$)[%] | 23.5 | 29 | 23.3 |

The HPA structure equipped with a beam loading similar to DDS_A, improves the rf-to-beam efficiency by ~5.5%. Reducing the beam loading degrades the efficiency back to the DDS_A. However, it reduces the surface fields significantly.

The other major concern is operating these structures is the wakefields induced by the beam. Another advantage of reducing the beam loading is related to the relaxed tolerance on the allowed transverse wakefield. The overall effect of the phase change on the dipole mode properties of each individual band (up to 6 bands) is discussed in the next section.

## DIPOLE BAND PARTITIONING

To rapidly compute the effect of increasing the phase advance on the first six dipole modes we utilise a 2-D mode matching code TRANSVRS [8]. Using TRANSVRS we calculate the first six dipole mode synchronous frequencies and kick factors for both the standard and HPA structure. As it is a 2-D code we cannot employ the manifold geometry. In addition, it is necessary to approximate all parts of the geometry with sharp steps, as the code cannot accommodate smooth surfaces. Nonetheless, the general features of the band structure in the wakefield are expected to be preserved. The kick factors are particularly sensitive to the aperture radius. Hence we do not expect the curvature in the geometry to affect the results significantly. In a structure of 24 cells, we choose seven cells, and obtain the remaining cell kicks and frequencies using spline interpolation fits.

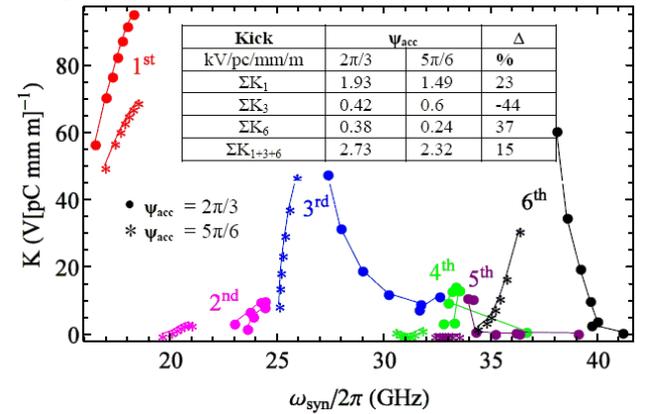

Figure 2: Comparison of kick factors of first six dipole modes in LPA and HPA detuned structure.

The kicks as a function of synchronous frequencies for the first six dipole bands are illustrated in Fig. 2 for both structure types. The 1st dipole band has the largest kicks. The next larger kicks are situated in the 3rd and the 6th bands. The summation of the 24 synchronous kicks in the 1st, 3rd and 6th band is 15% smaller in HPA compared to standard. Hence, the overall kick experienced by a beam in this structure is smaller than the standard structure. Suppression of the first dipole wakefield in a DDS_HPA structure is discussed in the next section.

## LOWEST DIPOLE MODE SUPPRESSION

The coupled mode wakefield in a manifold damped structure is calculated using a circuit model [2]. The multi-cell structure is first characterised in terms of parameters appropriate to a uniform structure, or a cell subjected to infinite boundary conditions. As an example of the procedure the circuit model applied to the two lowest HOMs (manifold mode and 1st dipole mode) in the first, middle and last single cell of a uniform structure are presented in Fig. 3. The degree of coupling of the attached manifolds to the cell modes is indicated by the avoided crossing in the coupled manifold-dipole curves.

Due to a strong coupling it is necessary to employ a Spectral function method [2] to calculate the coupling of the dipole modes to the manifolds. The Spectral function for two distributions is illustrated in Fig. 4 together with the symmetrical undamped distribution. In this case an eight fold interleaving is employed to add additional sampling of the distribution to satisfy the beam dynamics constraint.

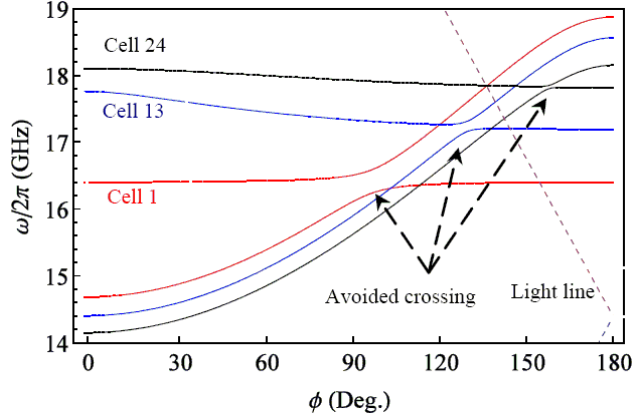

Figure 3: Dispersion curves of selected fiducial cells from a single periodic DDS_HPA structure. The degree of manifold-structure coupling is indicated by the degree to which the modes avoid crossing each other.

It is noticeable that in the HPA design the synchronous frequency of the first dipole mode shifts away from π phase advance. This results in an enhancement of the group velocity of the dipole mode by a factor of two (as it is ~ - 0.012c).

In order to facilitate sufficient coupling between manifold and dipole modes, the thickness of the irises has been modified. This has resulted in a reduced overall dipole mode bandwidth from 1.99 GHz to 1.44 GHz. The mode separation is largest at the ends of the frequency distribution and the kick factor is largest towards the upper frequency end. This is reflected in the difficulty of

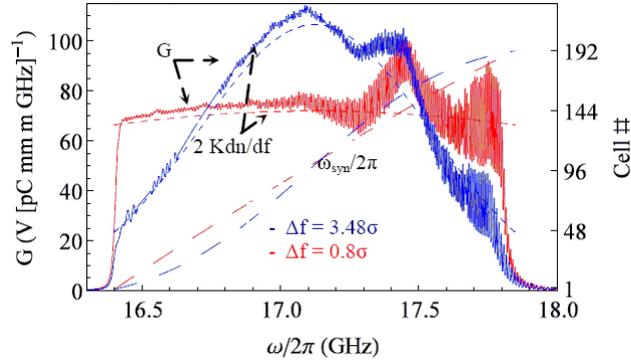

Figure 4: Spectral function (G) of an 8-fold interleaved DDS_HPA structure. The synchronous frequency, together with twice the kick factor weighted density function (2Kdn/df), are also shown for each distribution.

obtaining optimal coupling over the complete frequency range. There is the potential for additional optimisation in the upper frequency end of the distribution

The resulting wakefield, for a Gaussian distribution with a bandwidth similar to that of CLIC_DDS_A (Δf = 3.48 σ), illustrated in Fig. 5 (in blue), is now inadequately damped at the first trailing bunch. However, modifying the standard deviation of the Gaussian distribution, until it is now effectively almost a rectangular distribution, corresponding to a bandwidth of Δf = 0.8σ, accelerates the damping for the first few trailing bunches in particular (Fig. 5 red curve). In addition, the bunch population is reduced to $n_b$ = 3.2 x $10^9$. Both modifications ensure the beam dynamics constraints are adhered to.

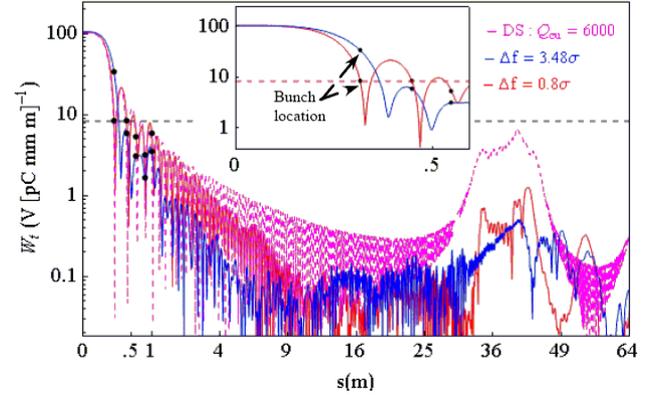

Figure 5: Envelope of wakefield of an 8-fold interleaved DDS_HPA structure, for two Gaussian distributions, over a complete train of 312 bunches. Also indicated is the envelope of the wakefield, for Δf =3.48σ, with the coupling to manifolds removed. Shown inset is the wakefield on the first three bunches.

## FINAL REMARKS

The motivation to investigate the 5π/6 structure was to reduce the accelerating mode's group velocity, whilst optimising the remaining parameters. In particular, the overall efficiency has been maximised, whilst the surface fields have been minimised and the long-range wakefields have been suppressed. Several of these parameters are superior compared to the DDS_A structure. In particular the required input power is reduced from 71 MW to 64 MW. The resulting efficiency is comparable to DDS_A but there is the potential for further optimisation.

## ACKNOWLEDGEMENTS

We are pleased to acknowledge useful discussions on the HPA structure with Z. Li and J. Wang of SLAC National Accelerator Laboratory and T. Higo of KEK. Research leading to these results has received funding from European commission under the FP7 research infrastructure grant no. 227579.